# Random Combinatorial Libraries and Automated Nanoindentation for High-Throughput Structural Materials Discovery

Vivek Chawla[1*], Dayakar Penumadu[1], Sergei Kalinin[2]

[1] Department of Civil and Environmental Engineering, University of Tennessee, Knoxville, TN 37996, USA
[2] Department of Materials Science and Engineering, University of Tennessee, Knoxville, TN 37996, USA

**Abstract**

Accelerating the discovery of structural materials is essential for advancing applications in hard and refractory alloys, hypersonic platforms, nuclear systems, and other extreme-environment technologies. However, progress is often constrained by slow synthesis–characterization cycles and the need for extensive mechanical testing across large compositional spaces. Here, we propose a rapid screening strategy based on random materials libraries, where thousands of distinct compositions are embedded within a single specimen, mapped by EDS, and subsequently characterized. Using nanoindentation as a representative case, we show that such libraries enable dense composition–property mapping while *exponentially reducing* the number of samples needed to span high-dimensional composition spaces compared to traditional synthesis-and-test workflows. An experimentally calibrated Monte Carlo framework is developed to quantify practical limits, capturing minimal particle sizes for reliable measurement, EDS noise and resolution, positional accuracy, and nanoindenter motion and reconfiguration costs. The simulations reveal regimes where random libraries provide orders-of-magnitude acceleration over classical workflows. Finally, we demonstrate the full experimental realization using macro-to-micro-EDS alignment, cluster-aware site selection, and automated indentation guided by cost-aware Gaussian-process planning. Together, these results establish random libraries as a general, instrument-agnostic route to high-throughput characterization in structurally critical material systems.

## I. Introduction

Material discovery is a cornerstone of innovation, underpinning industries worth trillions of dollars globally. From next-generation energy storage to structural alloys for aerospace, the search for materials with tailored properties drives technological progress. Yet, despite decades of advances, conventional trial-and-error experimental workflows remain painfully slow. Typical development cycles span years or even decades, constrained by manual "make-then-measure" loops that cannot keep pace with the urgency of modern demands.

Over the past two decades, compositional approaches, particularly high-throughput density functional theory (DFT) screening[1–4], have transformed early-stage exploration of material spaces. This theory driven pipelines enable rapid prediction of fundamental properties, guiding experimentalists towards promising regions of compositional and structural space. However, despite their predictive power, these approaches remain tethered to experimental validation. Predicted materials must still be synthesized and tested, meaning that the overall discovery cycle remains dominated by experimental bottlenecks.

To address this, automated synthesis platforms have emerged including robotic combinatorial deposition techniques[5,6], high throughput solution syntheses and autonomous flow reactors[7,8]. These advances allow researchers to generate thousands of unique compositions within a single experimental campaign. Yet, a persistent bottleneck of characterization remains[9,10]. Characterization throughput has not kept pace with syntheses or computation. Commercial characterization spanning X-ray diffraction, spectroscopy, and mechanical testing have experienced throughput improvement in decades. As a result, the bottleneck has shifted from synthesis to characterization, stalling the realization of truly autonomous "make-test-learn" discovery workflows. Closing this gap requires re-engineering existing tools into high-throughput characterization platforms capable of rapidly measuring complex, multidimensional material libraries.

[*] Corresponding author. Tel: +1 315-800-8228 E-mail address: vchawla@vols.utk.edu

Here we propose a fundamentally different strategy for accelerating characterization-limited discovery: random combinatorial libraries coupled with automated nanoindentation and cost-aware experimental planning[11]. In this approach, a single specimen encodes hundreds to thousands of discrete compositions through the stochastic spatial distribution of microscale particles, each representing an independent point in composition space. High-resolution energy-dispersive X-ray spectroscopy (EDS) provides chemical localization, while automated nanoindentation offers rapid, localized mechanical interrogation. Together, these elements transform a single physical specimen into a dense, high-dimensional composition–property dataset, bypassing the slow, serial constraints of traditional synthesis-plus-testing workflows.

We analyze this architecture from first principles, combining experimental measurements with a comprehensive Monte Carlo framework that captures realistic instrumental constraints, including minimum particle sizes for quantitative indentation, EDS noise and resolution, stage-motion penalties, and reconfiguration overheads. By integrating these experimentally calibrated factors into a unified active-learning model, we establish the theoretical and practical throughput limits of random libraries and benchmark their performance against conventional synthesis-nanomechanical workflows. This analysis reveals regimes where random library exploration yields orders-of-magnitude improvements in discovery rate and identifies key parameters-such as block size, positional accuracy, and noise tolerance-that govern the efficiency of high-dimensional search.

Finally, we operationalize the approach experimentally by constructing a particle-based random library, mapping its chemistry via EDS, and performing fully automated nanoindentation guided by Gaussian-process surrogate modeling. The resulting workflow demonstrates that random combinatorial architectures, paired with automated characterization and data-efficient planning, create a viable path toward high-throughput structural materials discovery and offer a practical route to closing the long-standing characterization bottleneck in autonomous materials research.

## II.     Background
### II.1   Bayesian Optimization with Gaussian Process
Bayesian Optimization[12–14] (BO) is a probabilistic framework designed to efficiently explore or optimize functions that are expensive to evaluate, often referred to as black-box functions. In this study, BO is employed to explore the unknown spatial variation of hardness across the XY domain of the simulation. The goal is to efficiently identify the hardness variation while minimizing the number of indentation measurements. At the core of BO is often a surrogate model that approximates the unknown black box function. In this work, gaussian process[14,15] (GP) is used as the surrogate due to its ability to provide both predictions and associated uncertainty estimates. A GP models the function as a multivariate normal distribution over all inputs (Equation 1). The radial basis function (RBF) kernel is used in this study given by Equation 2.

$$f(x) \sim GP(\mu(x), k(x, x')) \quad \text{-(1)}$$

$$k(x_i, x_j) = \exp\left(-\frac{\|x_i - x_j\|^2}{2l^2}\right) \quad \text{-(2)}$$

Here, $\mu(x)$ is the mean function (typically assumed to be zero), and $k(x, x')$ is the kernel or covariance function. $l$ is the kernel length scale, controlling how quickly the function varies across space. A small $l$ implies rapid variation or local noise, while a large $l$ indicates smooth, slowly varying behavior. During GP training, both the kernel parameters and the mean function are optimized to represent the observed data. The resulting model yields both a predictive mean $\mu(x)$ and the standard deviation $\sigma(x)$, enabling quantification of uncertainty at any point in the domain. Given the uncertainty and predictive mean, to guide sampling, BO uses an acquisition function which ranks candidate input locations based on their expected utility. Two acquisition functions used in this work are Upper confidence bound (UCB) and Uncertainty Estimation (UE) given by Equation 3 and 4 respectively.

$$UCB(x) = \mu(x) + \beta \cdot \sigma(x) \quad \text{-(3)}$$
$$\text{where, } \beta \text{ controls the trade off between exploitation and exploration}$$

$$UE(x) = \sigma(x) \quad \text{-(4)}$$
$$\text{prioritizes regions of highes model uncertainity}$$

### III. Random structural libraries

A random particle-based combinatorial library is a materials architecture in which discrete microparticles with independently assigned compositions are distributed stochastically across a substrate. Each particle occupies a finite region and contains a single, uniform composition drawn from a high-dimensional design space. Neighboring particles are compositionally uncorrelated, producing a spatial mosaic that embeds hundreds to thousands of distinct chemistries within one specimen. Because each particle is a separate composition, localized probes such as EDS or nanoindentation can interrogate a large compositional space without fabricating separate samples or relying on engineered gradients.

Several established combinatorial approaches exist for generating compositionally varied materials, including composition-spread thin films[16–18], multiple diffusion couples[19], mega libraries by C. Mirkin[20,21], and compositionally graded additive manufacturing[22]. These methods have been widely used to accelerate materials exploration and have enabled systematic mapping of low-dimensional chemistries. Although these methods accelerate exploration, they share common sampling limitations. Compositions evolve smoothly along one or two spatial axes, restricting the accessible dimensionality and leading to strong correlation between neighboring points. The number of distinct chemistries per sample is therefore modest, and the library remains anchored to a small set of endmembers or deposition targets. High-dimensional spaces, involving many simultaneous alloying elements or arbitrary multi-element mixtures, cannot be efficiently encoded, and acquisition of large discrete composition sets requires multiple samples or multiple deposition runs.

Random particle libraries bypass these constraints by assigning independent compositions to discrete particles distributed across a substrate. Because each particle is drawn without reference to specific endmembers or predefined mixing pathways, the library can incorporate a broad range of materials, including systems that cannot be co-synthesized or co-deposited using conventional methods. This architecture enables sampling of high-dimensional composition spaces within a single specimen, with each measurement probing a chemically unique region rather than a point along a gradient. As a result, random libraries can embed hundreds to thousands of distinct compositions, offering substantially greater diversity and coverage than traditional combinatorial methods and making them especially well-suited for automated, localized characterization workflows.

Figure 1 shows the contrast between classical gradient libraries and random particle-based libraries. Figure 1a shows a gradient library in which composition ($w_i^A, w_i^B \text{ or } w_i^C$) varies smoothly along spatial coordinates, whereas Figure 1b shows a random library where the elemental fractions ($w_i^A, w_i^B \text{ or } w_i^C$) at each particle are assigned independently of position. Although the arrangement in Figure 1b appears disordered in physical space, the corresponding property values in Figure 1c form a structured manifold in concentration space while remaining random in real space. Figures 1d and 1e illustrate surrogate model predictions and associated uncertainties over this landscape. Figure 1f shows an example acquisition function defined in both composition and physical space, and Figures 1g and 1h demonstrate how cost of measurement is incorporated to modify the acquisition function map and the decision is made based on the modified map (Figure 1h: note the decision for new measurement location is different compared to the one suggested by only acquisition function in Figure 1f).

Despite these advantages, random libraries introduce constraints that must be considered. Particle sizes are inherently variable, which creates nonuniform probe volumes and can push some particles below the spatial resolution of techniques such as nanoindentation or EDS. The stochastic spatial arrangement eliminates smooth compositional trends, making surrogate modeling more challenging and increasing sensitivity to measurement noise and positional error. Sharp compositional boundaries can amplify misalignment effects, and the lack of controlled synthesis pathways means that some particles may represent metastable or processing-dependent compositions that are difficult to reproduce. Together, these factors define the practical limits of how effectively random libraries can be interrogated and modeled, and motivate a systematic evaluation of their performance under realistic experimental constraints.

The current study takes a step forward this vision by analyzing the performance of a particle based combinatorial library model combined with active learning. Unlike traditional thin-film combinatorial libraries, which vary composition smoothly along one or two spatial axes, this model envisions a random dispersion of particles across a substrate. Each particle represents a unique composition drawn from a high-dimensional material space, enabling a single experimental campaign simultaneously characterize a vast range of chemistries. Post experiments, energy-dispersive spectroscopy (EDS) mapping and fiducial image registration provide the necessary chemical and spatial localization to assign compositions to mechanical or spectroscopic measurements. This architecture dramatically amplifies throughput. For example, a single indentation experiment can measure mechanical properties across hundreds or thousands of distinct compositions without sequential sample preparation. If successfully realized experimentally, the proposed random particle dispersion approach could increase high-throughput characterization rates by more than two orders of magnitude compared to conventional sequential workflows.

While promising, this approach introduces new challenges in data interpretation and optimization. Each measurement must be selected adaptively to maximize information gain while minimizing total experimental cost. Bayesian optimization (BO), with Gaussian Process (GP) surrogate models has been used in the literature and in this work as the framework for this task. BO navigates the trade-off between exploration and exploitation to accelerate discovery. However, the performance of GP-based active learning depends sensitively on experimental noise, spatial resolution, and the physical structure of the particle library itself. However, here we run BO with random spatial encodings – meaning that points with close concentrations can be very far from each other in real space.

This leads to several practical compromises. Increasing the number of particles expands the accessible composition space, but higher dimensionality introduces the usual sampling challenges and reduces the efficiency of Bayesian optimization. Instrumental factors also impose constraints, including the minimum particle size required for reliable mechanical measurements, stage-motion cost functions, and the probing volume of nanoindentation. On the SEM and EDS side, measurement time, elemental sensitivities, and spatial resolution further limit the fidelity with which compositions can be assigned to individual particles. To understand these dependencies, Monte Carlo simulations are conducted and the effect of different sources of experimental noise are implemented in the simulations.

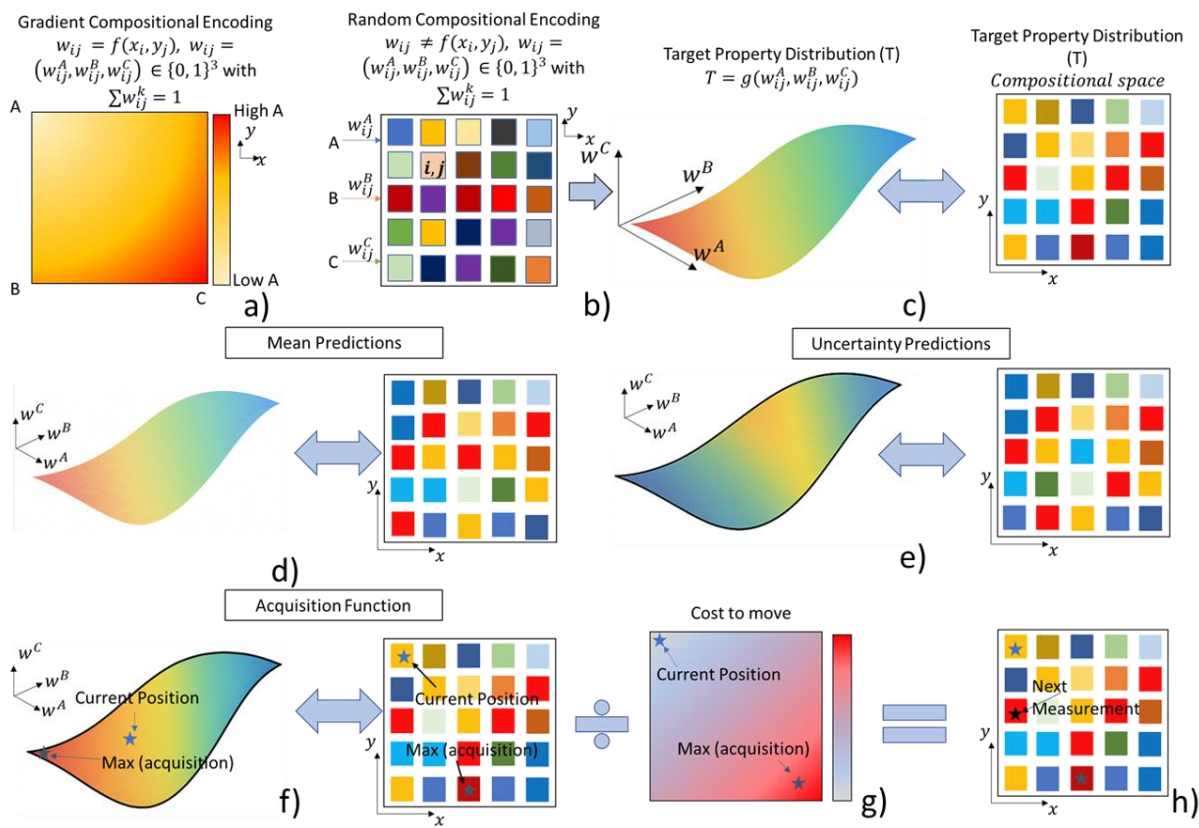

**Figure 1**: (a) Gradient libraries encode composition along spatial axes, (b) random libraries assign compositions independently of x and y, (c) these compositions form a smooth manifold in concentration space, (d–e) surrogate interrogation provides mean and uncertainty in both spaces, (f) these fields define a policy-based acquisition function, (g) motion constraints are introduced through a cost map, and (h) dividing acquisition by cost yields the final decision surface for choosing the next measurement.

## III. Random Library Simulations

### III.1  Simulated Random Library

A random particle-based combinatorial library is modeled as a spatial mosaic of microscale regions, each assigned an independent composition drawn from a high-dimensional design space. The substrate is represented as a square domain that is divided into nonoverlapping blocks whose sizes are sampled from prescribed minimum and maximum bounds. Each block is treated as a single particle with a uniform composition. The number of active elements and their fractions are chosen at random using a symmetric Dirichlet distribution. This construction produces a heterogeneous compositional field in which neighboring particles are uncorrelated and the overall domain samples a broad suite of chemistries.

To approximate experimental characterization, the ground truth composition field is down sampled to an effective EDS resolution and perturbed with Gaussian noise before renormalization. The resulting coarse gridded composition maps serve as the inputs for clustering, Gaussian process modeling, and active learning. Full details of the block generation procedure, Dirichlet sampling, and EDS simulation including particle tiling, domain sizes, and noise models are provided in the Supplementary Materials.

### III.1.1 Ground Truth

In this work, hardness is used as the ground truth property to align with the experimental focus on nanoindentation based characterization, although the qualitative conclusions extend to any high throughput probe. Hardness is a practical choice because composition dependent physics naturally produce complex and multimodal response surfaces. The same simulation framework could be applied to other indentation derived quantities, such as elastic modulus[23,24], hardness to modulus ratios, yield strength[25], ultimate tensile strength[25], fracture toughness[26,27] or creep resistance[28], but hardness is selected as the primary signal for clarity.

Ground truth hardness values are generated using physically motivated, composition dependent functions (Equation 5): rule of mixtures, entropy-based terms and periodic terms expressed through sinusoidal variation; full expressions are provided in the supplemental information. Figure 2a shows an example of the distribution of the hardness for a case of 6 elements (A-F) with varying composition. As dimensionality increases, these functions become increasingly intricate, yielding the complex landscape. For the Monte Carlo simulations, a simplified version of the hardness equations is used, also provided in the supplemental information. This simplification is necessary because highly complex formulations caused the surrogate regressor to degrade, making the resulting discovery trajectories increasingly random (supplemental information).

$$H(x, y, z) = f_{ROM}(x, y, z) + f_{entropy}(x, y, z) + f_{periodic}(x, y, z) \qquad \text{-(5)}$$

### III.1.2 Experimental Cost

A fully automated nanoindentation workflow[29] was implemented on a KLA iMicro system to enable high-throughput mechanical characterization across combinatorial libraries. The procedure follows a structured sequence involving sample initialization, optical targeting, grid definition, and automated execution of indentation cycles. Each cycle consists of Z-stage approach, surface engagement, drift stabilization, indentation under continuous stiffness mode, and subsequent withdrawal. The automation layer orchestrates X–Y motion between locations, manages optical focus, and monitors system state to ensure repeatable operation across large, spatially distributed grids. This enables integration with data-driven sampling strategies commonly used in combinatorial materials exploration.

To quantify experimental throughput, a cost model was constructed that captures the major time-dependent operations in the workflow, as described in our previous work. Figures 2b and 2c summarize this cost structure. Figure 2b shows the measured movement time as a function of distance, highlighting the nonlinear increase in cost for larger motions. In addition to movement cost, reconfiguration time is a significant contributor, as shown in Figure 2c. This reconfiguration cost is imposed by the manufacturer and is triggered when the distance between two consecutive indents exceeds approximately 1000 µm. In such cases, the stage cannot translate laterally in a single motion; instead, the z-stage fully retracts, the x–y stage moves to the new location, and the z-stage then re-engages and waits for drift stabilization, adding roughly 400 seconds of penalty. In the Monte Carlo simulations, we enforce movement constraints so that inter-indent distances remain below this reconfiguration threshold unless the penalty is intentionally modeled. This ensures that the simulated sampling behavior reflects the actual operational limits of the instrument.

Based on our previous work, a nominal hold time of 5 seconds is assumed for each experimental indentation. Using the measured timing contributions from drift, approach, withdrawal, and measurement, the effective duration for acquiring a complete 5 × 5 grid without triggering reconfiguration is approximately 1280 seconds. More details regarding these timing components and the derivation of this value can be found in our earlier study.

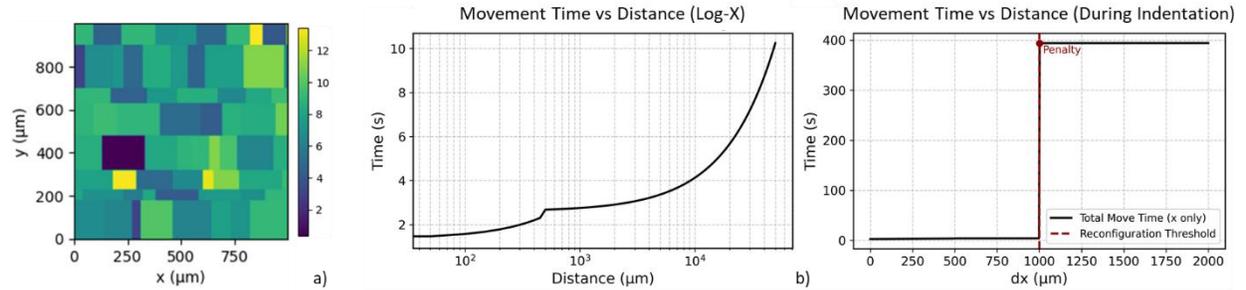

**Figure 2:** (a) Random library with 100 µm domain size and spatially varying hardness. (b) movement cost function of the XY stage. (c) cost function for movements within a grid, including reconfiguration cost.

### III.2 Encoded Information in Random Library

To understand the information gained using random library, hardness is assumed to be perfectly measurable at each sampled location. To assess the efficiency of GP modeling, the domain size is systematically varied while the size of the particle was kept constant (1unit * 1 unit), and the GP-predicted mean hardness is compared against the true governing function. Figures 3a and 3b illustrate two representative cases in which the particle size is fixed at one unit by one unit, but the overall domain size is varied. Increasing the domain from $50 \times 50$ units to $1000 \times 1000$ units increases the number of particles from 2,500 to one million, producing a much denser sampling of the concentration space. Although the underlying hardness function is identical in both cases, the achievable interpolation fidelity depends strongly on the domain-to-particle size ratio.

Two key aspects are examined as the number of compositional dimensions increases from 3D to 6D, where the highest-dimensional case also includes subspaces spanning 3D to 6D interactions. First, the mean absolute percentage error (MAPE) is evaluated. The purpose of this analysis is not to assess whether the Gaussian process can fully recover underlying governing equation (a regressor can be changed depending upon problem and domain knowledge), but rather to determine the domain size-to-particle size ratio required for the model to achieve convergence. Figure 3c presents the MAPE as a function of domain size with different dimensions. The results indicate that GP performance degrades as dimensionality increases (perhaps due to the discontinuous and complex governing equations). The convergence behavior stabilizes for domain size above $50 * 50\ units^2$.

Second, the minimum domain-to-particle size ratio needed to reliably detect the global maximum hardness ("best particle") is evaluated. Figure 3d shows the predicted maximum hardness as a function of domain size for 3D through 6D systems. Each point represents the average across thirty randomized seeds, with error bars indicating one standard deviation. As expected, the required domain size increases with dimensionality. Statistically, a domain-to-particle size ratio of approximately 200 is adequate for reliably identifying the global maximum in high-dimensional optimization problems. This threshold reflects the intrinsic sampling cost of exploring multicomponent compositional landscapes using a random particle-based library.

The number of independent compositional variables in an $N$-component system is $N - 1$, since the elemental fractions must sum to unity. A ternary system therefore requires only two independent coordinates, which can be represented fully within a single two-dimensional library. However, as dimensionality increases, direct visualization and exhaustive sampling become increasingly impractical. If each independent variable is discretized into eleven evenly spaced composition intervals (10% interval), the total number of distinct sub-libraries required to span the full composition space grows as shown in Equation 6. For a quaternary system this yields 11 distinct libraries, and for a six-component system, 1331 libraries. This exponential growth illustrates why traditional gradient or spread-film approaches cannot scale to high-dimensional chemistries and motivates the need for architectures such as random particle libraries, which embed very large numbers of unique compositions within a single specimen.

$$Number\ of\ libraries = 11^{n-3} \qquad \text{-(6)}$$

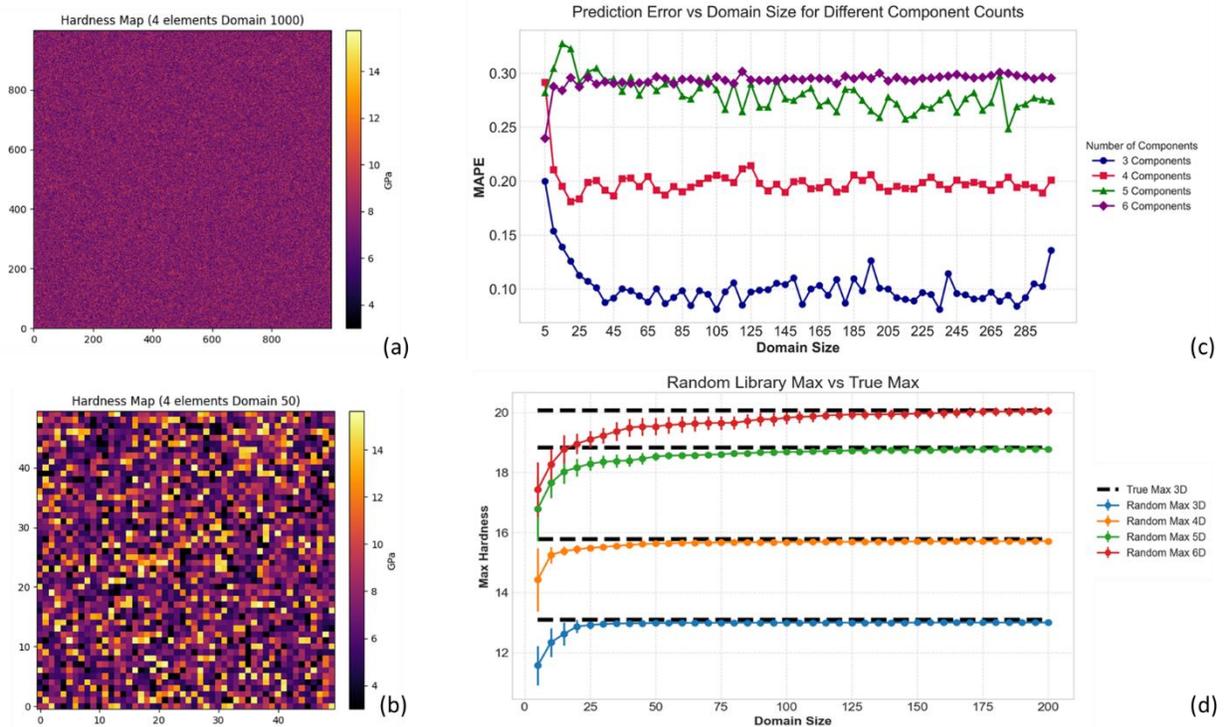

**Figure 3:** (a, b) hardness maps over domains of size 1000 units and 50 unit, respectively, using the same particle size of 1 * 1 unit$^2$. (c, d) performance of a vanilla GP in capturing the underlying physics and locating the maximum hardness, shown as a function of domain size and the number of elements sin the composition.

## III.3 Monte Carlo simulations

*III.3.1 Simulation Framework*

At a high level, each Monte Carlo run draws a random experimental scenario and then simulates a full cost-aware active learning experiment on that virtual sample. First, the domain size ($L$), block size range ($a_{min}, a_{max}$), number of blocks, EDS pitch ($r_{EDS}$), EDS noise ($\sigma_{EDS}$) , and positional error ($\sigma_{XY}$) are sampled from prescribed distributions (details on range is shown in Figure 4 top). Given these parameters and a random seed, a particle-based library is generated by tiling the domain with blocks, assigning each block a random multicomponent composition, and evaluating the governing hardness model to obtain the "true" hardness field in both physical and compositional space. Note that once the library is created the true domain variables as well as the number of particles ($N$) can also be recorded. An EDS map is then simulated by adding Gaussian noise with standard deviation ($\sigma_{EDS}$) to the true compositions on a coarse EDS grid and renormalizing the concentrations. Because the EDS map is pixelated (defined by EDS pitch), compositions at arbitrary indentation locations are obtained by bilinear interpolation of this noisy map. The details of the whole workflow can be found in supplemental information.

Using the noisy, interpolated compositions as inputs, a Gaussian Process surrogate is fit in composition space and used to drive a cost-aware Bayesian optimization loop. At each global iteration, the surrogate predicts mean hardness and uncertainty across candidate regions, combines them with the time cost of moving the stage and performing indents, and selects the next location to probe. Around each chosen center, a fixed 5*5 grid of indentation points is planned. Before executing these points, a positional offset is applied, drawn from a normal distribution with zero mean and standard deviation ($\sigma_{XY}$), to emulate stage positioning error. Indentations are then simulated at the shifted locations by sampling hardness with measurement noise, and the surrogate is updated using the *actual* indentation location, reflecting the fact that the post-indentation imprint reveals the location of indentation. The run terminates either when any visited point lies sufficiently close to a true hardness hotspot and attains near-maximum hardness, or when 100 global iterations have been completed. Figure 4 provides the whole workflow.

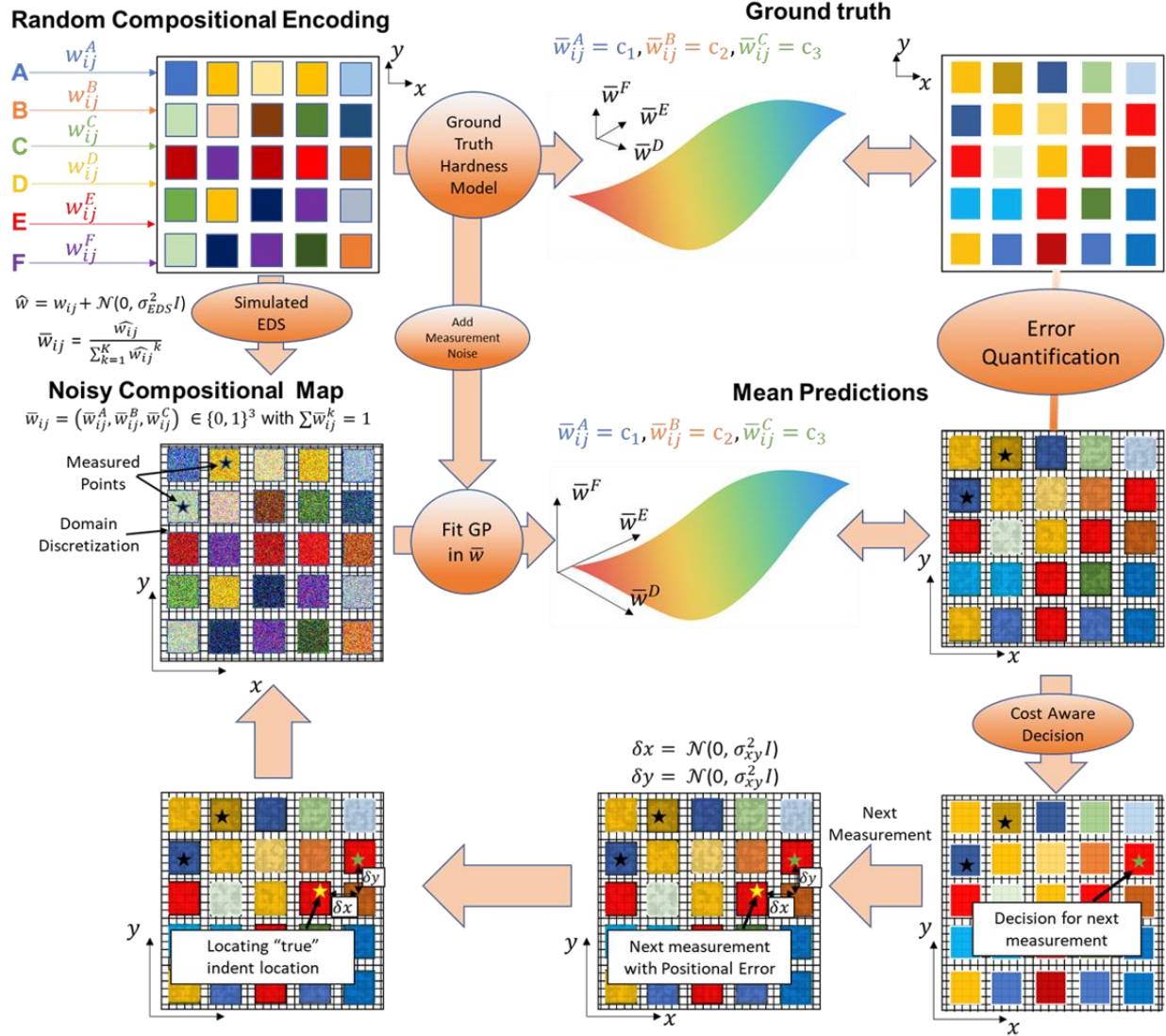

**Figure 4**: Workflow for the Monte Carlo simulation of random particle libraries, including parameter sampling, ground-truth generation, noisy EDS reconstruction, GP fitting, cost-aware acquisition, and indentation with positional error.

### III.3.2 Model

After generating the full set of Monte Carlo simulations, symbolic regression is applied to extract generalizable relationships between discovery time and the underlying structural and experimental parameters. The analysis begins by forming engineered variables: $\frac{\sigma_{XY}}{a_{min}}$, $\frac{\sigma_{XY}}{a_{max}}$, $N_{blocks}$, $\sigma_{EDS}$, $r_{EDS}$. These features and the discovery iteration count are normalized, and a symbolic regression model is trained using PySR. The framework produces a Pareto front that spans models of increasing complexity and accuracy. From this set, the simplest expression that maintains stable predictive performance is selected. All PySR search settings, operator definitions, and additional results are provided in the supplementary materials.

*III.3.3  Results*

The Monte Carlo simulations reveal clear patterns that determine when discovery succeeds and how difficult the search becomes under experimentally realistic uncertainty. Figure 5a shows the distribution of iterations to discovery across all simulations. Many runs converge within a few tens of iterations, while a sharp peak at the 100-iteration limit corresponds to cases where the algorithm never sampled the true optimal region. Figure 5b summarizes pairwise correlations among all parameters. Iterations to discovery correlate most strongly with geometric features of the library, including the number of blocks and the minimum and maximum block sizes, and only weakly with surrogate absolute percentage error, positional error, or EDS noise. Interestingly, EDS noise is the only parameter that correlates meaningfully with surrogate error, reinforcing that chemical noise primarily affects model fidelity rather than search difficulty directly.

Figure 5c compares parameter distributions for successful and unsuccessful runs. The successful runs were defined as those in which the discovery was achieved within 100 iterations, corresponding to roughly 35 hours of experimental time. The most pronounced separation appears in the number of blocks and the minimum block size. Failed simulations concentrate at high block counts and very small block sizes, regimes that create a highly granular landscape with many spatially uncorrelated regions. In such conditions, the Gaussian process receives limited smooth structure to exploit, and the active-learning loop must explore numerous local optima before reaching the global maximum. Positional error also shows a systematic shift, with failures exhibiting larger $\sigma_{XY}$ values. When blocks are small, positional uncertainty causes the instrument to land off-center and probe mixtures of neighboring compositions, effectively missing the "optimal" particle. EDS noise and resolution show more modest differences, influencing discovery primarily through their impact on compositional accuracy and subsequently absolute percentage error (ape) rather than through direct changes in hardness variability.

Figure 5d compares the predicted and true iterations from symbolic regression and shows that, despite noticeable scatter, the analytical expression captures the dominant monotonic trends in search difficulty.

Because many simulations converged early, iteration counts were weighted according to their frequency so that heavily populated early-convergence cases did not dominate the regression. The large scatter arises not only from the number of variables in the system but also from the fact that many of these variables represent uncertainty rather than fixed values. For example, a positional noise level of $\sigma_{XY} = 5$ does not imply a constant error of 5µm. Instead, it indicates that the positional error is drawn from a normal distribution $\mathcal{N}(0, 5)$, which perturbs each location differently while maintaining an average scale of variability governed by the value 5. The engineered variables $A$, $B$, and $C$ (Equation 8-10) normalize positional uncertainty and particle granularity, allowing the learned expression (Equation 7) to generalize. The selected symbolic-regression equation reflects the same structure seen in the Figure 5b: terms involving C dominate. Failure to discover the "best" particle arises primarily when C is large or when A and B become large, indicating positional error is comparable to or larger than particle size, which significantly reduces the ability to strike the correct block (Figure 5c). Overall, the symbolic regression expressions provide compact summaries of these dependencies, showing that block density controls iterations to discovery, positional error determines the recoverable spatial scale, and EDS noise mainly affects surrogate fidelity rather than the fundamental ability to "discover". These results are encouraging because they show that once $\sigma_{xy}$ is known, the values of $a_{min}$, and $a_{max}$ can be chosen such that, even with coarse and noisy EDS maps, the automated experiments will still converge to the optimal material. The main consequence is simply that larger $N$ increases the total time required, but does not fundamentally prevent discovery.

$$I = 0.15 * \exp\sqrt{\sqrt{C} * [6.68 * (\exp(AB) * \sqrt{C * \exp(\sigma_{EDS})})]} \qquad \text{-(7)}$$

$$A = \frac{\sigma_{xy}}{a_{min}} \qquad \text{-(8)}$$

$$, B = \frac{\sigma_{xy}}{a_{max}} \qquad \text{-(9)}$$

$$, C = N_{Blocks} \qquad \text{-(10)}$$

Where, $I$ is the iterations to discovery, $\eta$ is EDS noise, $m$ is the minimum block size, $N$ is the number of blocks, and x is the positional error, and $r$ is the EDS resolution

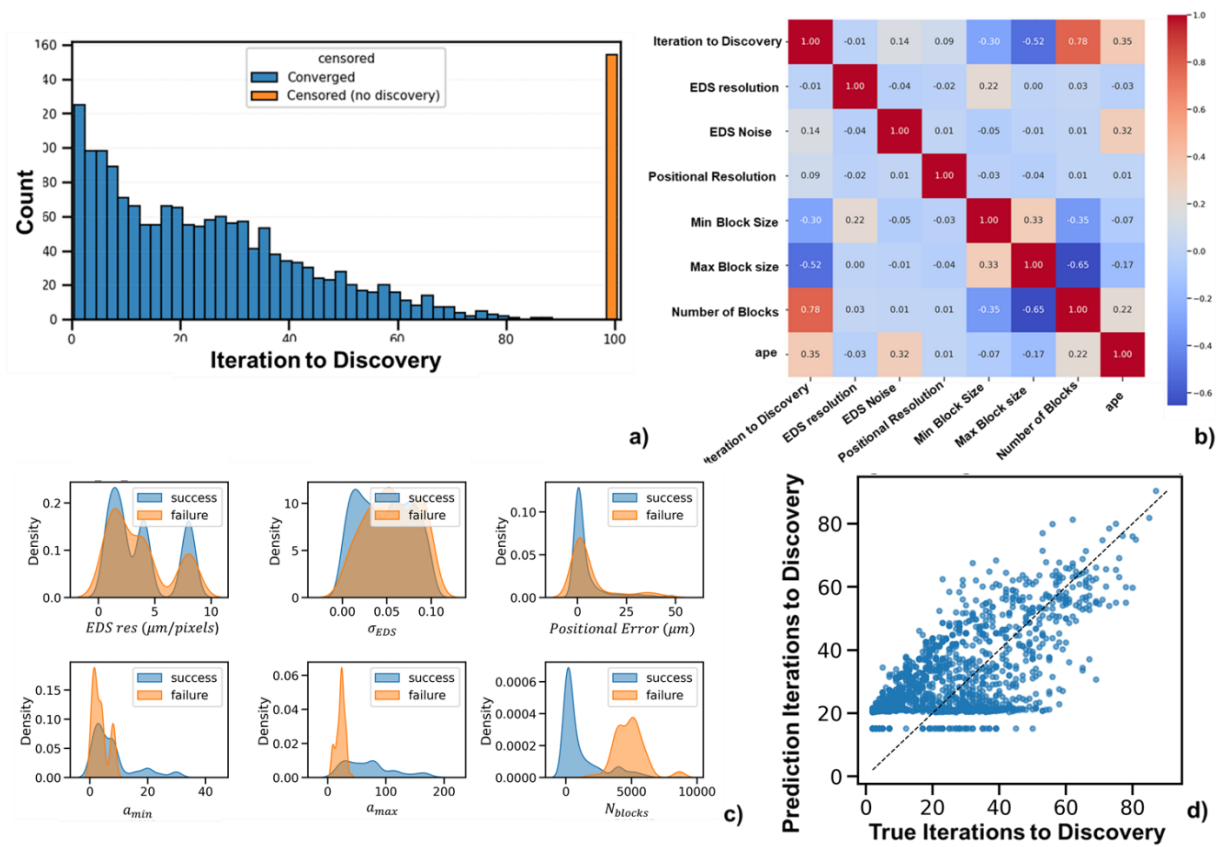

**Figure 5**: (a) Iteration distribution showing early convergence and the 100 iteration non-converged peak. (b) Correlations among all features, with strongest dependence on block size and block density. (c) KDEs contrasting successful and unsuccessful runs, highlighting the effects of block count, block size, and positional error. (d) Parity between predicted and true iterations from symbolic regression.

## IV.    Experimental realization

An important advantage of the random library approach is that it is compatible with a wide range of materials systems and sample preparation routes. It does not depend on specific synthesis methods, endmembers, or deposition conditions, and can therefore be realized using virtually any source of particulate matter. To demonstrate this compatibility, we prepared and measured an experimental random library using heterogeneous metallic particulates and characterized their local chemistry by EDS.

### IV.1   Experimental Random library and EDS characterization

Here, A random materials library was prepared using metallic particulates obtained as residual chafing from machining operations. The particulates, representing diverse alloy systems with uncontrolled chemistries, were randomly selected and embedded within a brass ring using conductive epoxy. The epoxy served as the mounting medium yielding a consolidated puck in which metallic inclusions were stochastically dispersed throughout an insulating matrix. Therefore, the sample consists of small number of discrete particles per library, but the heterogeneous origins of the particulates lead to a combinatorically rich and effectively high-dimensional chemical space.

Energy-dispersive X-ray spectroscopy (EDS) mapping was then performed to establish the local chemical composition at each spatial location in the library. EDS was performed using a scanning electron microscope (SEM, Helios 5 DualBeam, Thermo Fisher) equipped with an energy-dispersive X-ray spectrometer (EDXS, EDAX Octane Elite, AMETEK) and an electron backscatter diffraction detector (EBSD, EDAX Velocity, AMETEK). Imaging was performed using the in-chamber electron detector (ICE). Two representative areas were analyzed at an accelerating voltage of 15 kV and a current of 6.4nA with a working distance of 4 mm. The dead time was ~ 40, and the total count rate were maintained at ~150,000 counts/s. Each map consisted of 2048 * 1600 pixels and was acquired until a maximum count for any element ~30 was reached, corresponding to ~20-30 minutes of acquisition. Two complete mapping iterations were conducted under identical conditions. Quantification was performed using ZAF quantification provided in the vendor software (EDAX apex SOFTWARE).

The resulting hyperspectral data were exported in HDF5 (.h5) format, with a full EDS spectrum stored at each pixel. Subsequent analysis was carried out using the pyTEM[30] and HyperSpy[31] Python libraries. To improve computational tractability and suppress pixel-level noise, the datasets were spatially down sampled by a factor of 8, yielding effective images of 200 * 256 pixels. Pixel-wise quantification was conducted using ±50ev integration windows centered on kα emission lines of twelve selected elements. Due to the limited dwell time and variable background intensity across the epoxy- metal interfaces, the resulting dataset constitutes a deliberately noisy case of EDS quantification. Figures 6a and 6b display the total X-ray counts for the original and down sampled datasets for an area within the random library, respectively. Phase segmentation based on the EDS hyperspectral counts was performed using nonnegative matrix factorization, and the full analysis is provided in the Supplemental Information.

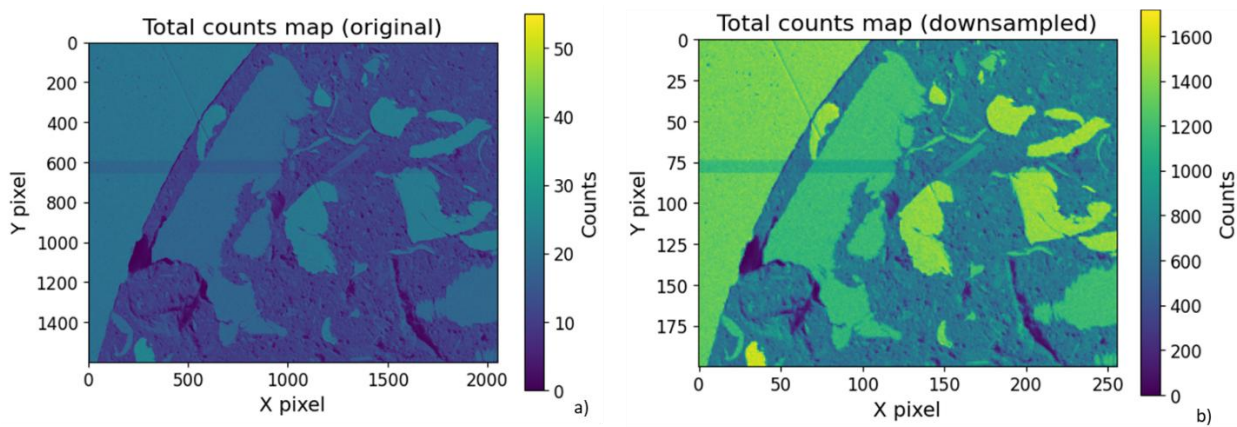

**Figure 6:** (a, b) EDS colormaps of total X-ray counts for Area 1 (original and down sampled).

### III.2. Automated Nanoindentation

Automation navigation is then performed to enable targeted indentation of these particles. The workflow follows the macro-to-micro alignment procedure described in previous work. In this process, the downsampled and reprocessed EDS maps (with C and O removed) are exported and used as the "macro" reference images. During alignment, two reference points are visited and the system computes the rotation and translation needed to register the EDS map to the true X–Y orientation of the indenter stage. Once aligned, pixel coordinates from the EDS map can be used directly for stage navigation.

After alignment, valid indentation sites are identified. The library contains particulates dominated by steel, copper, and brass, and naïve random sampling would bias indentation toward these abundant classes. To obtain representative coverage, unsupervised K-means clustering is applied to the quantified EDS data across all valid pixels. The optimal number of clusters, determined by silhouette analysis, is n = 7. Candidate indentation centers within each cluster are selected using two geometric constraints: each site must be at least 5 µm from the particle boundary, and centers must be spaced by at least 50 µm to prevent interference between grids. After applying these constraints, small particulates such as aluminum yield only a few valid sites, whereas larger particles support up to 100 sites. The details of the library construction and selection procedure are provided in the Supplemental Information and in Figure SI3.

Once the alignment and site selection are complete, a Gaussian process (GP) model is used to guide the indentation campaign. Although the composition space is high dimensional and noisy, initial exploration is carried out using three randomly selected 3×3 indentation grids spaced at 5 µm. Indents are performed to a target depth of 200 nm with a Poisson ratio of 0.3 and strain rate of 0.2, assuming total depth is equal to contact depth. Modulus values below 10 GPa are removed to exclude epoxy regions, and the grid-level modulus is taken as the median of the remaining measurements. If fewer than three valid indents remain, the grid modulus is assigned as NaN.

Figure 7 shows the two mapped areas rotated to match the nanoindenter coordinate frame. Figures 7a–d show the GP-predicted mean and uncertainty for both areas. As expected, the uncertainty remains high due to the limited number of measurements and the high dimensionality of the compositional space. Figure 7c marks the current location and the next chosen indentation site, with an inset showing the corresponding optical view from the indenter.

An exploratory phase of ten indentation grids is performed using an acquisition parameter $\beta = 2$ to emphasize uncertainty-driven exploration. Because the compositional dimensionality is on the order of ten and the dataset is sparse, the GP is not expected to accurately capture the modulus relationship (Section III.2). To account for discontinuities arising from distinct compositional families, the GP prior uses a seven-segment piecewise linear mean function, matching the number of clusters. During the exploratory phase, the GP samples broadly across compositional classes and obtains representative modulus values.

Site selection at each iteration is determined by maximizing the acquisition function normalized by combined travel and measurement costs. Since movement cost is negligible relative to the ~five-second indentation time per location, the measurement cost dominates the normalization. After the exploratory phase, $\beta$ is reduced to 1 to balance exploration and exploitation. A final stage is performed with $\beta = 0.1$, temporarily increased to 1 every third iteration to reintroduce targeted exploration. Figures 7e and 7f show the resulting sampling sequence, with early and late measurements distinguished by color, while Figure 7g plots the discovered modulus as a function of iteration. In this case the maximum modulus was identified early, but the staged transition from exploration to focused search demonstrates the robustness of the overall strategy. Occasional NaN values correspond to grids located near particle edges.

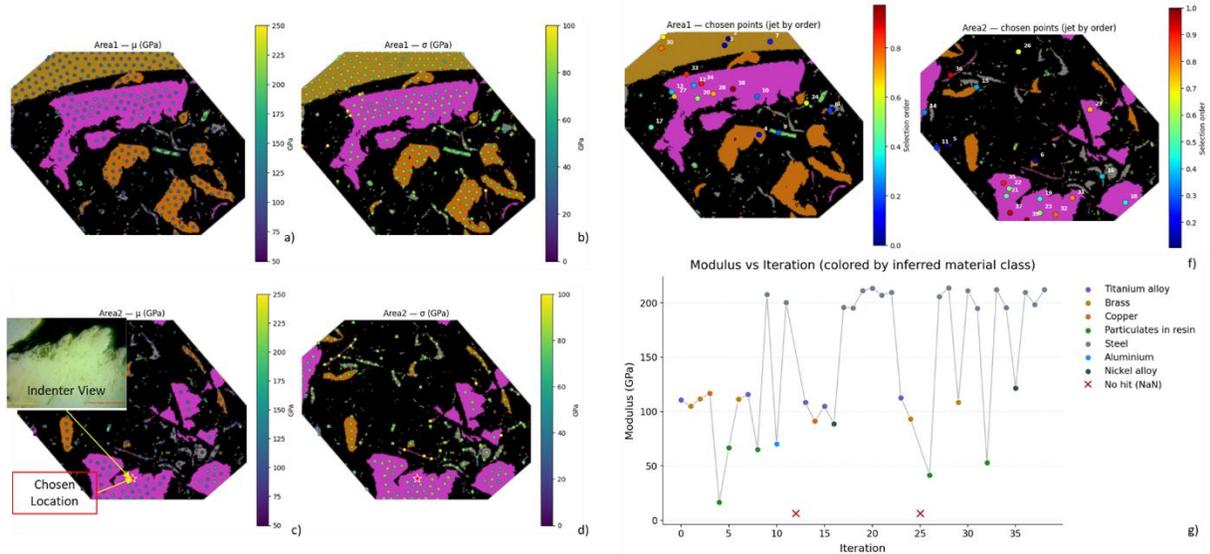

Figure 7. (a-d) GP-predicted mean and uncertainty for two areas; in (c) the star marks the next selected location, and the inset shows the indenter view after arrival. (e, f) all visited indentation sites, with points colored by recency using the jet colormap (red = recent, blue = earlier). (h) acquisition history illustrating the shift from exploration to exploitation.

## V. Conclusions

This study introduces random particle-based combinatorial libraries as a pathway to accelerate high-throughput materials discovery. Unlike conventional thin-film or gradient-based libraries, which encode compositional variation along one or two spatial axes, the random library architecture disperses a large number of discrete particles across a substrate, each representing a distinct point in a high-dimensional compositional space. This design substantially increases the amount of encoded information available per experiment, enabling the simultaneous exploration of complex chemistries that would otherwise require many sequential libraries. When coupled with automated characterization platforms such as nanoindentation or spectroscopy, the random library provides an orders-of-magnitude improvement in throughput, since each measurement contributes not only to the property of a single composition but also to mapping across a broad compositional domain.

The simulations reveal that domain-to-particle size ratio and block density are the primary geometric controls on discovery difficulty. A ratio of roughly 200 is sufficient to reliably locate the global optimum even in six-dimensional composition spaces. Symbolic regression shows that the normalized block density term $C = \frac{N}{L^2}$ dominates iterations to discovery, while normalized positional-error terms $A = \frac{\sigma_{XY}}{a_{min}}$ and $B = \frac{\sigma_{XY}}{a_{max}}$ act as secondary modifiers. EDS noise influences surrogate fidelity but has limited impact on the fundamental ability to converge as long as each particle contains at least one measurable point. These results indicate that once $\sigma_{xy}$ is known, particle dimensions can be chosen so that automated experiments remain robust even with coarse, noisy EDS maps; higher particle counts simply increase the time, not the probability of failure.

Experimentally, we show that the entire workflow, from macro-to-micro-EDS alignment to cluster-aware indentation planning and cost-aware acquisition, can be deployed on a commercial nanoindenter. The same architecture naturally extends to processing-space exploration using inert markers, to construction of synthesis pipelines that feed randomized composition fields, and to multi-fidelity schemes where nanoindentation serves as a rapid proxy for mechanical testing. Together, these elements position random combinatorial libraries as a practical and scalable approach for accelerating structural materials discovery.

**Acknowledgements**
This research was primarily supported by National Science Foundation Materials Research Science and Engineering Center (MRSEC) program through the UT Knoxville Center for Advanced Materials and Manufacturing (DMR-2309083).

## SUPPLEMENTAL INFORMATION

### SI. Simulated Random Library

A random combinatorial particle library is conceptualized as an idealized framework to evaluate the throughput and precision of automated characterization workflows. Instead of limiting each substrate to a single, uniform composition, a dense dispersion of microscale particles is assumed, where each particle represents a unique point in the compositional design space. The configuration facilitates parallel exploration of a high-dimensional composition space within a single experimental setup. Such particle-based libraries may arise from various experimental pathways. These include mixing of pre-alloyed powders, combinatorial syntheses techniques like co-sputtering, or process-induced heterogeneity as seen in severe plastic deformation methods such as high-pressure torsion (HPT). These techniques can generate local chemical gradients and stochastic compositional variations across the microstructure. Although real systems are influenced by processing conditions and thermodynamic constraints, the present study assumes a purely random and uncorrelated distribution of particle compositions across the substrate. The objective is to evaluate the limits of high throughput exploration in an idealized setting.

To simulate a random library, a square physical domain of side $L$ (ranging between 200-1000 $\mu m$) is selected as the substrate for each simulation. This domain is tessellated into non-overlapping rectangular blocks, each representing a particle with uniform local composition. The block dimensions are randomized by drawing side lengths from a uniform distribution between user-specified minimum and maximum limits (that varies between simulations) and truncated at the boundaries to guarantee complete spatial coverage without gaps. Tiling proceeds band-by-band from top to bottom and left to right, producing a random mosaic of particles whose sizes varied between simulations but always conformed to the domain boundaries. Each block is then assigned an independent composition vector: the number of active elements is chosen randomly between one and the global dimensionality $D$; the corresponding elements are selected without replacement, and their fractions are sampled from a symmetric Dirichlet distribution, then normalized to unity. This procedure results in particle compositions varying from single-element blocks to mixtures spanning the full global dimensionality $D$. The result is a spatially resolved ground-truth composition field, $c_{true}(x, y)$, which is constant within each block but heterogeneous at the scale of the overall domain thereby resulting in a "random particles" distribution across a domain. As the number of particles within the domain increases, the system becomes more structurally complex; however, this enhanced complexity also enlarges the sampled design space, thereby increasing the likelihood of capturing a previously unexplored or "novel" composition with desirable properties.

To mimic real experimental scenarios, the composition field is determined using simulated energy-dispersive spectroscopy (EDS) mapping. The true composition is discretized onto a coarse grid where the spacing is set by the effective resolution $r$ ($\mu m$ per pixel). In practice, this means that the instrument returns one averaged composition vector for each $r \times r$ $\mu m$ patch. Gaussian noise with iteration specific standard deviation is added independently to each channel of the coarse grid, and the perturbed vectors are renormalized to enforce non-negativity and unit sum. This process yields the simulated EDS composition field, $c_{EDS}(i, j)$, at each pixel. An example random library is illustrated in Figure S1. The top row shows the true per-element composition maps (A–F) at full resolution across the $1000 \times 1000$ $\mu m$ domain. Each map is uniform within a particle and varies only at block boundaries, reflecting the random tiling and Dirichlet-sampled compositions. The bottom row shows the corresponding simulated EDS measurements. In this case, the EDS resolution is $r = 10$ $\mu m$ per pixel, so the 1000 μm domain is sampled on a $100 \times 100$ grid. The axes are therefore reported in EDS pixels rather than physical μm. The speckled appearance within blocks arises from the added Gaussian noise, which perturbs the otherwise uniform local compositions. Together, the two sets of maps demonstrate how random combinatorial domains are generated and how simulated measurement noise alters the apparent composition fields available for clustering, Gaussian process modeling, and active learning.

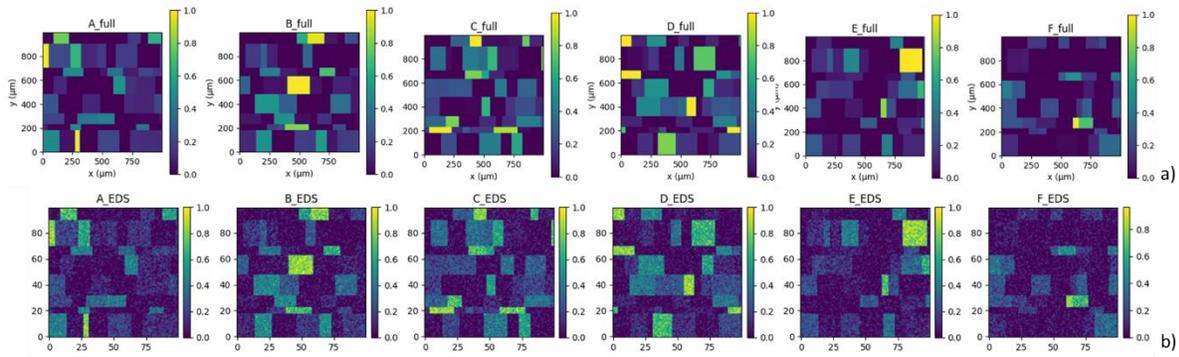

Figure S7: Example compositional map of a random library with varying particle size and elemental composition. (a) Ground truth compositions. (b) Simulated noisy concentration as would be measured by EDS

## SI. Experimental Cost

The experiments are performed using a KLA iMicro Nanoindenter. An automated nanoindentation framework is implemented. The nanoindentation procedure typically follows a standardized sequence. First, the sample is mounted on a holder and placed within the indenter. The holder moves from its safe position to a location beneath the optical lens using X-Y stage movement, and the Z-stage is adjusted to bring the surface into focus. Following initialization, indentation grids are defined. Once grid setup is complete, the indentation cycle begins. The lift step involves raising the Z-stage to a safe height and moving the X-Y stage to the first indent location. The engage step follows where the Z-stage lowers to identify the surface contact. After the engage step, the system undergoes drift stabilization to minimize drift before testing.

A comprehensive cost model is developed to account for the time associated with various operations in the nanoindentation workflow. These costs are categorized into two groups: fixed action-level overheads, and variable context-dependent penalties. Fixed operations include the lift step, engage step, and return step, which correspond to tip approach, surface contact, and withdrawals, respectively. Based on experimental timing, the lift step requires approximately 30 s, engage step takes 65s, and the return step takes an additional 35 s. In addition to the fixed costs, the variable time penalties include drift stabilization time, measurement duration, and sample movement time. Drift stabilization is assigned an average value of 300 s. While this value can vary in practice depending on environmental stability, it serves as an estimate for the present model. The measurement time of 120 s, consistent with CSM-based nanoindentation targeting a depth of 200 nm is set. This duration accounts for surface detection, loading, unloading, and drift monitoring phases (~80 s by default). Stage movement time is modeled as a function of the distance between consecutive indentation locations. Timing data for movements ranging from 2 µm to 50,000 µm were recorded and used to fit a piecewise function. At short distances, movement time follows an exponential trend due to PID-controlled motor dynamics, while long-range motion behaves linearly.

While performing an indentation within a grid, a critical behavioral shift occurs when the distance between indents exceeds 1000 µm. At this threshold, the instrument applies a reconfiguration penalty to prevent tip damage during long movements, a constraint given by the manufacturer. This involves lifting the indenter, performing lateral movement, re-engaging at the new location, and re-stabilizing drift. The cumulative penalty for this reconfiguration sequence is approximately 400 s. As a result, spatial sampling strategies must account for this discontinuity in cost when designing efficient indentation grids. The total time required to complete a single indentation grid is given by the sum of all the fixed and variable components: lift, engage, drift stabilization, measurement, return, and movement. The relative contribution of each component depends on the spatial arrangement of the indents. For closely spaced grids, drift stabilization dominates the overall cost structure, particularly in configurations with a small number of indents. As grid size increases, the absolute drift stabilization time remains constant, but its relative contribution to total testing time decreases. Consequently, the percentage of time spent on actual indentation measurements increases with grid size, improving overall time efficiency. For widely spaced grids (those crossing the reconfiguration threshold), movement and re-engagement become the dominant factors. Within the measurement step itself, a substantial portion of time is spent on hold for drift correction, which adds redundancy given the initial drift stabilization phase.

## SII. Full Workflow for the Monte Carlo Active-Learning Framework

Each Monte Carlo simulation begins by generating a randomized experimental scenario that mimics the variability encountered in real combinatorial characterization. The domain size, EDS step size, block size range, EDS noise, positional error, and coverage fraction are all sampled from predefined distributions that span realistic instrument conditions. The domain size is chosen from a discrete set and is adjusted so that it aligns with the EDS pixel grid. The minimum block size is drawn from a fixed set, and the maximum block size is computed from a randomly sampled ratio while enforcing both a lower bound (to ensure the block is larger than a single pixel) and an upper bound (to prevent blocks from exceeding half the domain). A target areal coverage is drawn from a uniform range, and the expected number of blocks is computed from the expected block area under a uniform distribution. The actual block count is then sampled from a normal distribution around this estimate, with the spread reflecting natural variability in particle packing. EDS noise is sampled uniformly over a moderate range, and positional error ($\sigma_{XY}$) is sampled from a log-uniform distribution covering sub-micron to tens-of-microns variability, matching the range observed in automated nanoindentation stages.

Given this parameter set, a random particle library is constructed by tiling the domain with rectangular blocks whose side lengths fall within the sampled bounds. Each block is treated as one microparticle. Its composition is drawn by selecting a random subset of the six available elements, assigning Dirichlet-distributed fractions, and normalizing them to sum to one. This produces a highly heterogeneous spatial mosaic where the composition of one particle is statistically independent of its neighbors. Once the full composition field is defined, the smooth 6-component hardness model is evaluated at each block, producing the ground-truth hardness distribution across the domain. This model incorporates linear mixing, quadratic cross-interactions, sinusoidal coupling, an equiatomic Gaussian bump, and an entropy-based term, enabling realistic non-monotonic and multi-modal structure in the high-dimensional composition space. The hardness field is then upsampled to a 1 µm grid so that indentation coordinates and motion costs can be simulated at micrometer resolution. All pixels whose hardness lies within a small tolerance of the global maximum are recorded as valid "discovery" targets.

To simulate EDS-based composition mapping, the true composition map is sampled on a coarse grid defined by the EDS pitch. Gaussian noise with the sampled standard deviation is added independently to each elemental channel, after which the vectors are renormalized back onto the simplex. Because indentation locations generally do not lie exactly on the EDS grid, compositions at arbitrary coordinates are obtained by bilinear interpolation of the noisy EDS map followed by an additional small Gaussian perturbation before final renormalization. This process captures both the discretization error imposed by finite measurement resolution and the uncertainty introduced by the EDS detector.

The surrogate model is initialized by selecting a small number of global seed locations from a sparse grid. Around each seed, a fixed 5×5 indentation pattern is planned with 5 µm spacing. Before hardness is queried, each indentation point is shifted by drawing a random positional offset from a normal distribution with standard deviation equal to the sampled positional error. This represents the experimental scenario in which the instrument does not land exactly on the requested coordinates. The true hardness at each offset location is then sampled with a small measurement noise term. The Gaussian Process surrogate is trained on these noisy compositions and hardness values, using separate GP models for the mean and noise components. To ensure numerical stability across thousands of Monte Carlo runs, the GP training set is capped at 250 points by random subsampling when necessary.

The cost-aware active learning loop proceeds iteratively. At each iteration, the surrogate model predicts mean hardness and uncertainty at every candidate global sampling region. These predictions are divided by the estimated time cost of moving the stage, engaging, stabilizing, and performing the indentation sequence. The acquisition score is therefore the expected measurement value weighted by the inverse of the motion and setup cost, ensuring that the algorithm avoids expensive long-range moves unless they provide substantial informative benefit. The region with the highest score is selected, and a new 5×5 indentation block is executed. As before, every indentation point is displaced by a positional jitter drawn from a normal distribution, and the actual landing coordinates—not the intended ones—are used when updating the surrogate model. This reflects experimental reality: the visible indent imprints provide the true measurement locations, and the GP must incorporate this information.

Depending on the surrogate's predictive uncertainty within the newly sampled region, the algorithm may also construct a refined local grid with higher spatial density. This adaptive refinement is controlled by estimating the variance of the surrogate on the initial 5×5 block; if the uncertainty is too large, the local grid is expanded to increase information density. After each global and local update, the GP is retrained using capped sample sizes, and both the mean and noise models are refreshed.

A discovery check is performed at the end of each global iteration. The algorithm compares all visited actual indentation locations against the set of true hotspot pixels. Discovery is declared only if a visited point lies within a fixed spatial tolerance of a hotspot and the measured hardness at that location is within the target fraction of the global maximum. If this does not occur, the simulation continues until either discovery is achieved or the algorithm reaches 100 iterations. Each Monte Carlo run records the evolution of prediction error, kernel length scales, accumulated cost, predicted maxima, visited points (ideal and actual), the full composition maps, the full hardness field, and statistical summaries of block geometry. All data are saved for downstream meta-analysis.

## V.1 Symbolic Regression

A detailed symbolic-regression analysis was performed on the full Monte Carlo dataset to extract generalizable relationships between discovery time and the underlying structural and experimental variables. After the simulations were completed, the raw parameters for each run were transformed into a set of engineered features designed to reflect the relevant physical scales: positional error normalized by both the minimum and maximum block sizes, block density normalized by the domain area, EDS resolution, and EDS noise. These features were min–max normalized, and the target variable (iterations to discovery) was normalized in the same manner. Censored cases (simulations that reached the iteration limit) and trivial low-iteration runs were excluded to ensure that the regression reflected meaningful convergence behavior. Symbolic regression was then performed using PySR, which generates candidate analytical expressions through an evolutionary search that balances predictive accuracy against algebraic complexity. The operator set consisted of addition, subtraction, multiplication, division, square, square-root, and exponential transforms, and a small parsimony penalty was applied so that increased expression complexity was only accepted if it produced a corresponding improvement in fit quality. PySR maintained a Pareto front of solutions with varying complexity, and models were ranked by their predictive score and symbolic complexity as determined by the internal cost function. From this Pareto front, the expression selected for interpretation was the lowest-complexity model whose predictive performance remained comparable to the top-performing (but more complex) formulas. The selected model was then mapped back to the original iteration units, and its performance was validated through parity plots, residual analysis, and one-dimensional response curves generated by sweeping each normalized feature while holding the others at their median values. Full details of operator definitions, complexity costs, PySR hyperparameters, and the complete equation table are provided in the supplementary notebook.

## SIII. EDS Phase analysis

To further interrogate the EDS hyperspectral data, we applied nonnegative matrix factorization to the down sampled EDS cube. The spectrum at each pixel was first reshaped into a two-dimensional matrix X of size (number of pixels) by (number of energy channels), and pixels with zero total counts were masked to avoid numerical issues. NMF was then used to factor X into two nonnegative matrices $X \approx W H$, where each row of H represents a characteristic spectrum and each column of W gives the corresponding abundance for every pixel. We used six components, so that W contains six abundance fields and H contains six associated basis spectra. After factorization the abundance coefficients in W were mapped back to image coordinates to generate spatial phase maps for each component. Figure S2 shows the resulting six NMF phase abundance maps. Each map emphasizes a different microstructural motif, including epoxy rich regions, fine matrix domains, and distinct inclusion families, demonstrating that unsupervised NMF on the noisy EDS dataset can separate chemically and morphologically distinct phases and provide a physically interpretable segmentation of the random particle library.

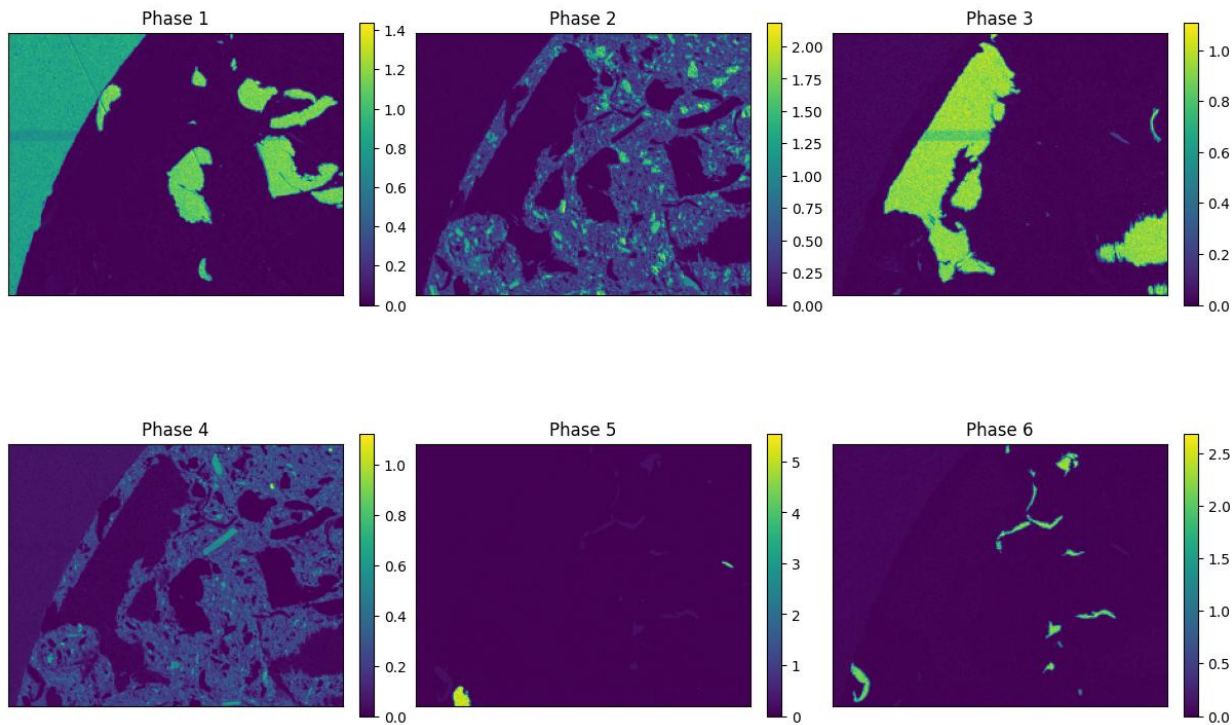

**Figure S8**: Nonnegative matrix factorization applied to the downsampled EDS hyperspectral cube, showing six extracted phase abundance maps. Each map corresponds to one NMF component and highlights spatially distinct chemical or microstructural regions within the random particle library.

## SIV. Experimental Random Library

Figure S3a presents an optical micrograph of the random library, while Figures S3b and S3c show composite EDS maps of two representative areas. Figure S3d shows the specific regions from which the EDS maps are acquired. To reduce noise and improve computational efficiency, the hyperspectral EDS data are first down sampled. Pixels in which the combined concentration of C and O exceeded 80% are excluded, as these corresponded to the epoxy-rich regions. Due to the limited conductivity of the epoxy and associated charging effects, the EDS maps are inherently noisy, which should be considered when interpreting the results. Figure S3e displays the two analyzed areas after removal of the high C + O regions (the black areas in the Figure), followed by re-quantification under the assumption of negligible carbon and oxygen content. This procedure isolates the metallic particulates although some residual contributions from the conductive epoxy filler remain. Figures S3f and S3g illustrate the rotated and aligned particle locations used for this navigation step. To achieve more representative coverage, unsupervised clustering (K-means) is applied to the quantified EDS data for all valid pixels. The number of clusters was determined by silhouette analysis, which yielded an optimal solution of n=7 clusters. The resulting cluster assignments are shown in Figures S3h and S3i for the two mapped areas. The selected indentation centers are shown in figures S3j and S3k for the two mapped areas.

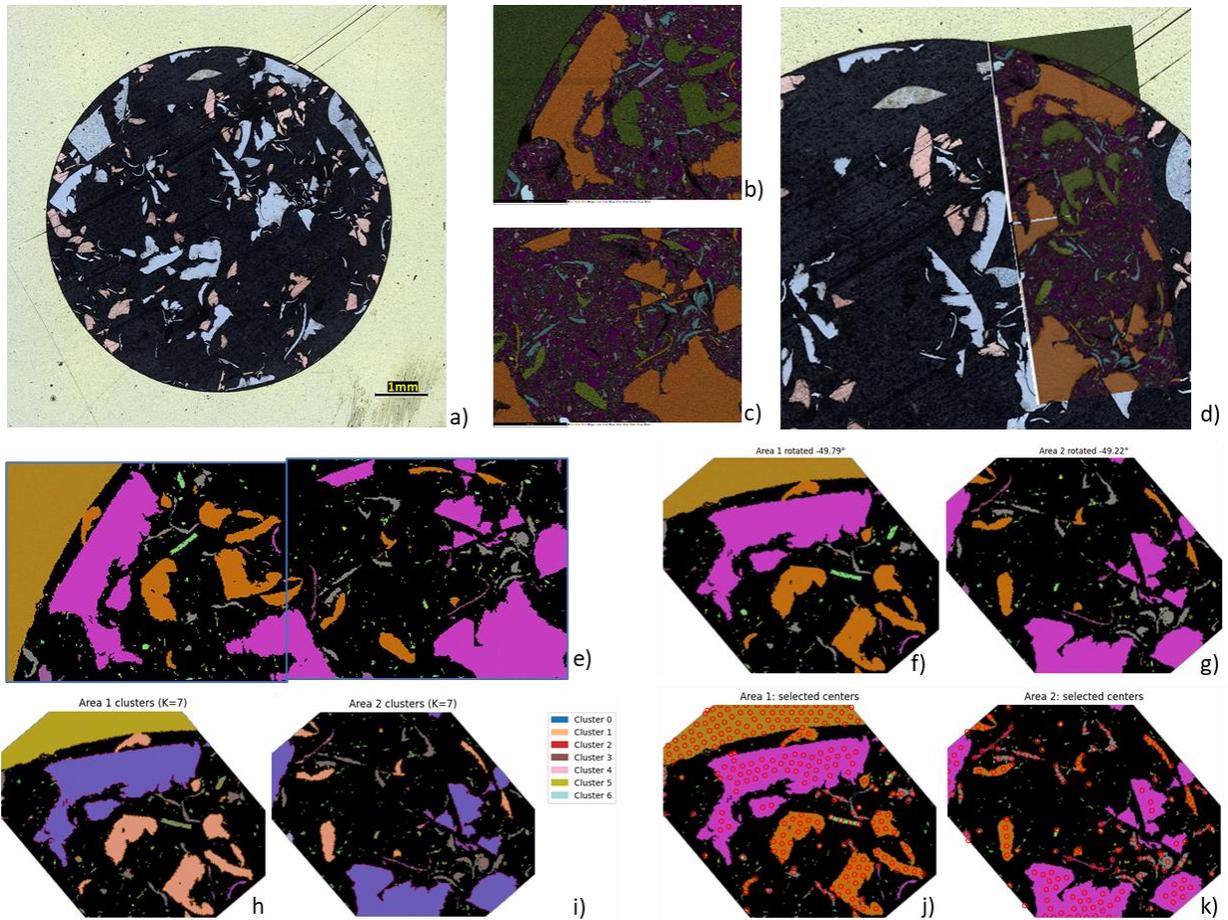

Figure S3: (a) Optical micrograph of the random library with particulates embedded in epoxy within a brass ring. (b, c) EDS maps for two areas. (d) Locations of the EDS acquisitions on the sample. (e) composition map after masking polymer dominated pixels. (f, g) automated alignment of the two EDS maps. (h, i) phase identification via k-means clustering with silhouette analysis for the two areas. (j, k) prospective indentation sites chosen randomly from clusters.